\documentclass[fleqn,twoside]{article}

\usepackage{espcrc2}
\usepackage{graphicx}
\usepackage{epsfig}
\usepackage{color}
\usepackage{graphicx}     
\usepackage{dcolumn}        
\usepackage{bm}         
\usepackage{amsmath}
\usepackage{amssymb}
\usepackage[figuresright]{rotating}

\title{Probing the Standard Model via rare pion and muon decays}

\author{E. Frle\v{z}\address{Department of Physics,
        University of Virginia, Charlottesville, VA~22904-4714, USA}
        \thanks{for the PIBETA Collaboration}}
       
\begin{document}

\begin{abstract}
The PIBETA collaboration has used a non-magnetic pure CsI calorimeter 
operating at the Paul Scherrer Institute to collect the world's largest 
sample of rare pion and muon decays. We have extracted the absolute 
$\pi^+\to \pi^0 e^+\nu$ decay branching ratio with the 0.55\,\% 
total uncertainty. The $\pi^+\to e^+\nu\gamma$ data set was used to extract
weak axial and vector form factors $F_A$ and $F_V$,
yielding a significant improvement in the precision of $F_A$ and $F_V$.
The $\mu^+\to e^+\nu\nu\gamma$ distributions were well
described with the two-parameter $(\rho_{\rm SM},\bar{\eta}=0$) solution.
These results bring major improvements in 
accuracy over the current Particle Data Group listings and agree 
well with the predictions of the Standard Model.
\vspace{1pc}
\end{abstract}

\maketitle

\section{INTRODUCTION}
\begin{figure}
\noindent \hglue 0.2cm (i)
\hbox{\ }\vglue -1.0cm
\hglue 0.5cm\includegraphics[width=6.0cm]{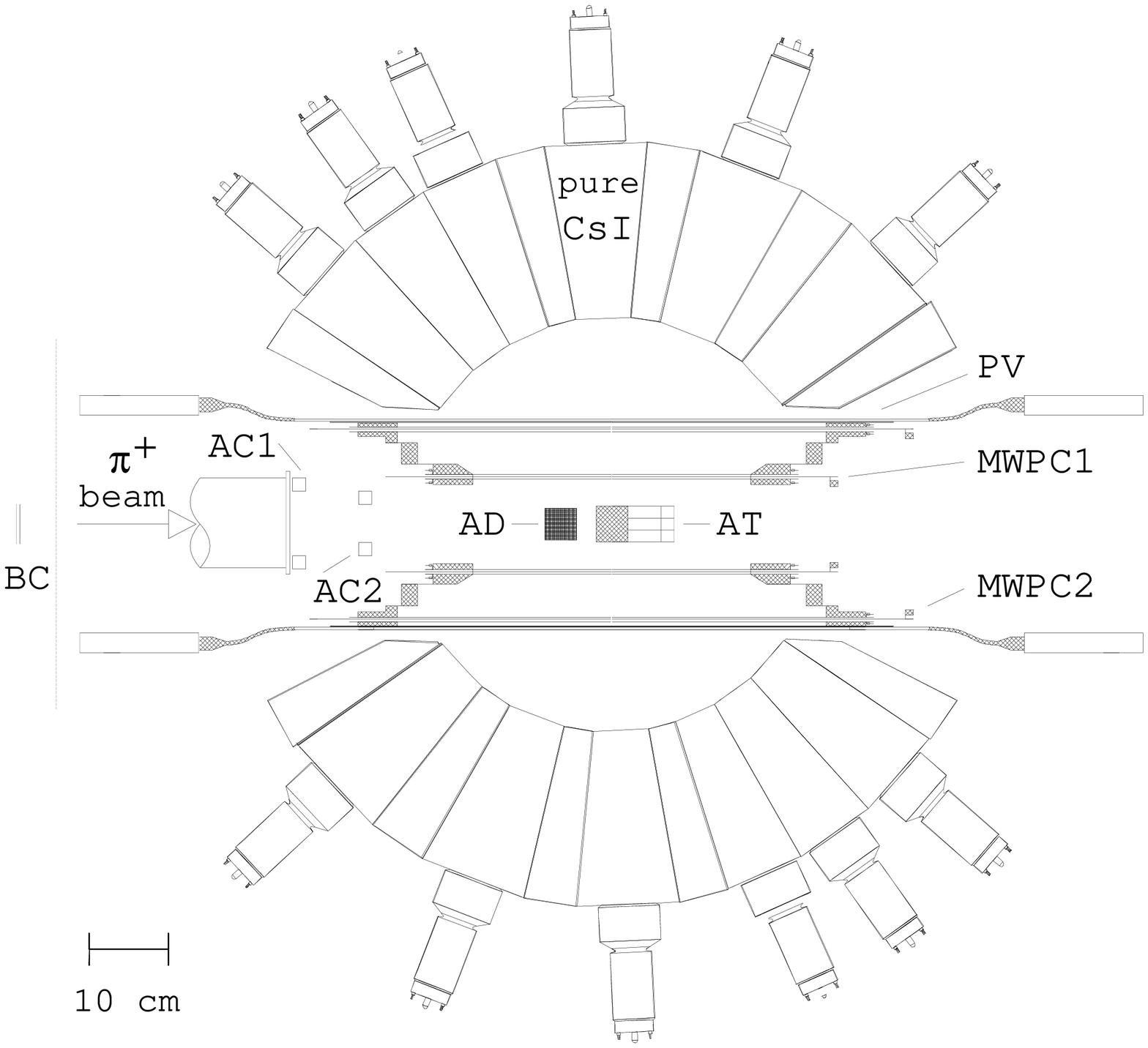}\\
\noindent \hglue 0.2cm (ii)
\hbox{\ }\vglue  -0.3cm
\hglue 1.6cm\includegraphics[width=4.2cm]{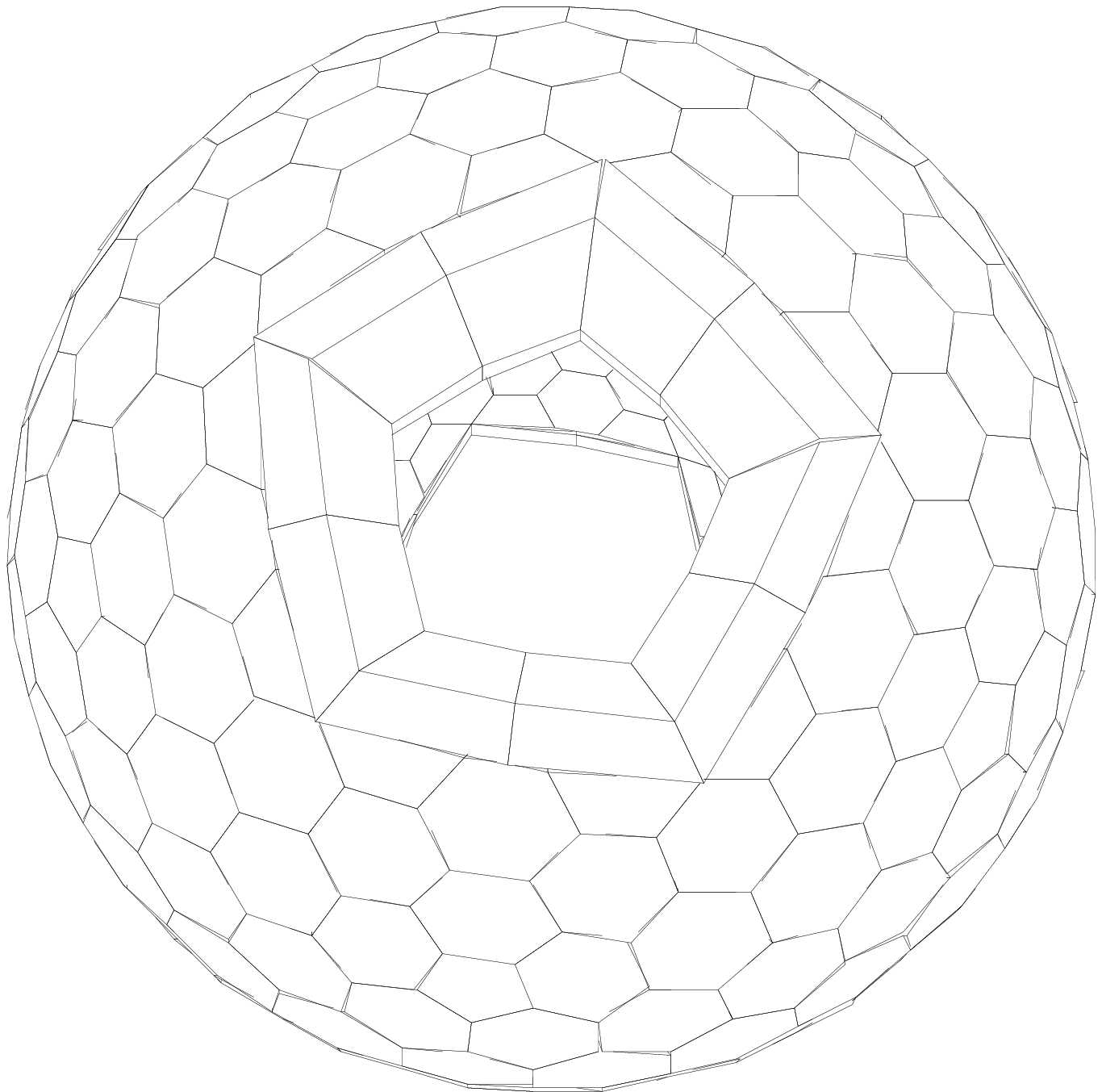}
\vglue -1.3cm
\caption{Pibeta detector: 
(i) cross-sectional view showing
the major detector sub-systems, and (ii) 240-module
pure CsI calorimeter geometry.}
\label{fig:det}
\end{figure}

The PIBETA experiment at the Paul Scherrer Institute (PSI) is 
a collaboration of seven institutions that collected the world largest 
sample of rare pion and muon decays during the 1999-2001 and 2004
beam periods~\cite{www}. 

The PIBETA detector system is based on a large acceptance
240-module pure CsI electromagnetic shower calorimeter. 
The detector includes
with an active degrader AD, a segmented active target AT, 
a 20-bar cylindrical plastic scintillator veto PV for 
particle identification, a pair of tracking cylindrical multi-wire 
proportional chambers MWPC1/2 and an active cosmic veto shield~\cite{Frl04a}.

A schematic drawing of the detector is shown in Fig.~\ref{fig:det}.
The incident 114\,MeV/c $\pi^+$ beam with a minimal
$e^+$/$\mu^+$ contamination was tagged with a thin forward beam
counter BC, slowed down in the degrader and ultimately stopped
in a tight $\sigma_{x,y}\simeq\,9\,$mm spot in the active target.
The 1999-2001 runs, optimized for the pion beta decay
measurement, used beam fluxes of up to 1\,MHz. Beam intensities 
of 50-200\,$\pi^+$/s were used in the 2004 for 
optimal acquisition of radiative decay events.

The recorded data, comprising $2.2\cdot 10^{13}$ $\pi^+$ 
stops, were obtained by a dedicated two-arm high-threshold trigger
as well as 11 physics and calibration triggers, some of which 
were prescaled.

\section{PION BETA DECAY}
\begin{figure}
\includegraphics[width=7.5cm]{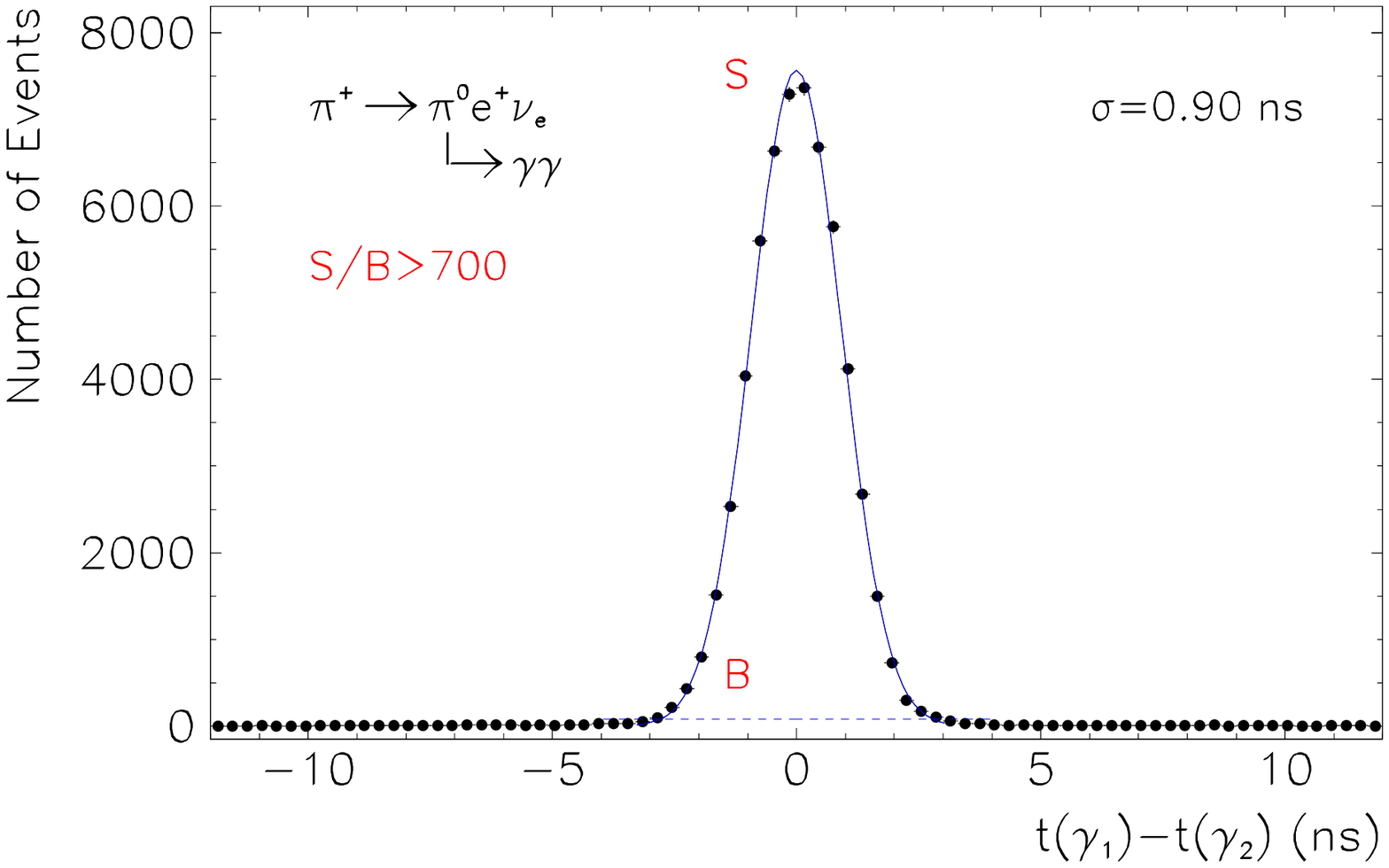}\\
\vglue -0.8cm
\includegraphics[width=7.5cm]{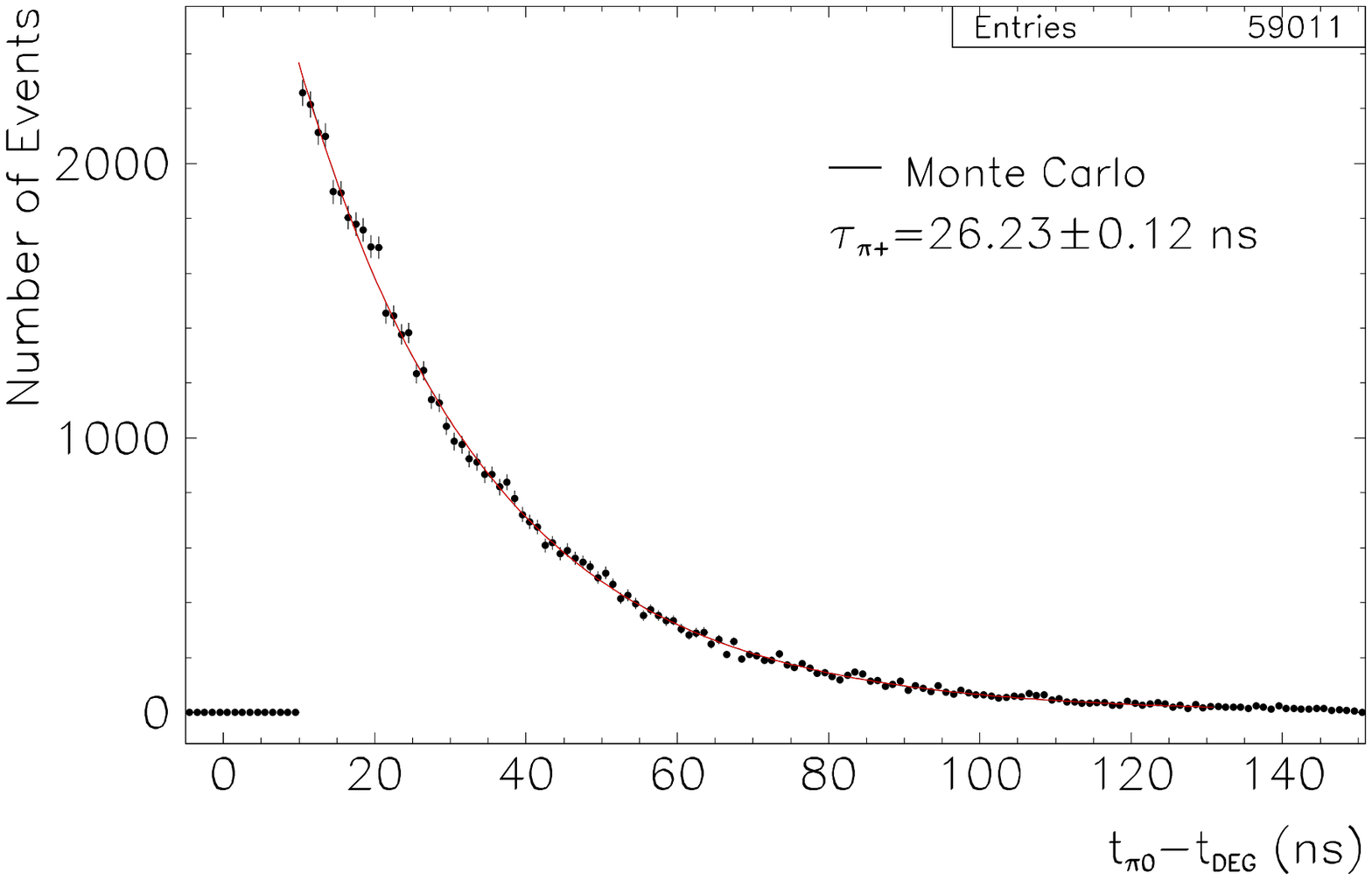}
\vglue 0.7cm
\caption{Representative pion beta decay experimental spectra: 
(i) Signal-to-background ratio, and
(ii) measured life time for the $\pi\beta$ events.}
\label{fig:pb}
\end{figure}

Pion beta decay ($\pi^+\to \pi^0 e^+\nu$) rate
offers one of the most precise means of testing the conserved
vector current hypothesis (CVC)~\cite{Fey58} and studying the weak $u$-$d$ quark 
mixing~\cite{Cab63}. The Standard Model (SM) description of the $\pi\beta$ decay   
is theoretically unambiguous within a 0.1\,\% uncertainty~\cite{Jau01,Cir03}, 
but a small $\sim 1\cdot 10^{-8}$ branching ratio poses significant 
experimental chalenges. The 3.9\,\% uncertainty of the previous most 
precise measurement, 
made using the $\pi^0$ spectrometer at LAMPF~\cite{McF85},
was not accurate enough to test
the full extent of radiative corrections which stand at 
$\sim 3$\,\%~\cite{Mar86}.

The fast analog hardware triggers were designed to accept nearly 
all non-prompt $\pi\beta$ and and a sample of prescaled $\pi^+\to e^+\nu$ 
events with individual shower energy exceeding 
the Michel endpoint (high threshold $\simeq$~52 MeV).

The data analysis provided clean distributions of 64,047 
$\pi\beta$ decay events which agreed very well with energy, angular 
and timing spectra predicted by the GEANT3 Monte Carlo detector
simulations. The cosmic muon, prompt, radiative pion and accidental
backgrounds were determined to be $<1/700$ of the $\pi\beta$ signal,
Fig.~\ref{fig:pb}.

We have chosen to normalize the $\pi\beta$ yield to the yield of
$\pi^+\to e^+\nu$ events whose branching ratio is known with 0.33\,\% 
uncertainty experimentally~\cite{Bri92,Cza93} and $\le 0.05$\,\% accuracy 
theoretically~\cite{Mar93,Dec95}. Using the PDG~\cite{Eid04} 
recommended value of $R^{\rm exp}_{\pi\to e\nu}=1.230(4)\cdot10^{-4}$,
we find the pion beta branching ratio~\cite{Poc04}:
\begin{equation}
\label{eq1}
R^{\rm exp}_{\pi\beta}=\left[ 1.036\pm 0.004{\rm (stat)}
\pm 0.005{\rm (syst)}\right] \cdot 10^{-8}.
\end{equation}
When normalizing to the theoretical value 
$R^{\rm the}_{\pi\to e\nu}=1.2353\cdot 10^{-4}$\cite{Mar93} we obtain:
\begin{equation}
\label{eq1p}
R^{\rm exp}_{\pi\beta}=\left[ 1.040\pm 0.004{\rm (stat)}
\pm 0.005{\rm (syst)}\right] \cdot 10^{-8}.
\end{equation}
Our result for $R^{\rm exp}_{\pi\beta}$ is in excellent agreement
with the prediction of the SM:
\begin{equation}
\label{eq2}
R^{\rm SM}_{\pi\beta}= (1.038-1.041) \cdot 10^{-8},
\end{equation}
and stands as the most accurate confirmation of the
CVC in a meson to date.

\section{RADIATIVE PION DECAY}
\begin{figure}
\includegraphics[width=7.5cm]{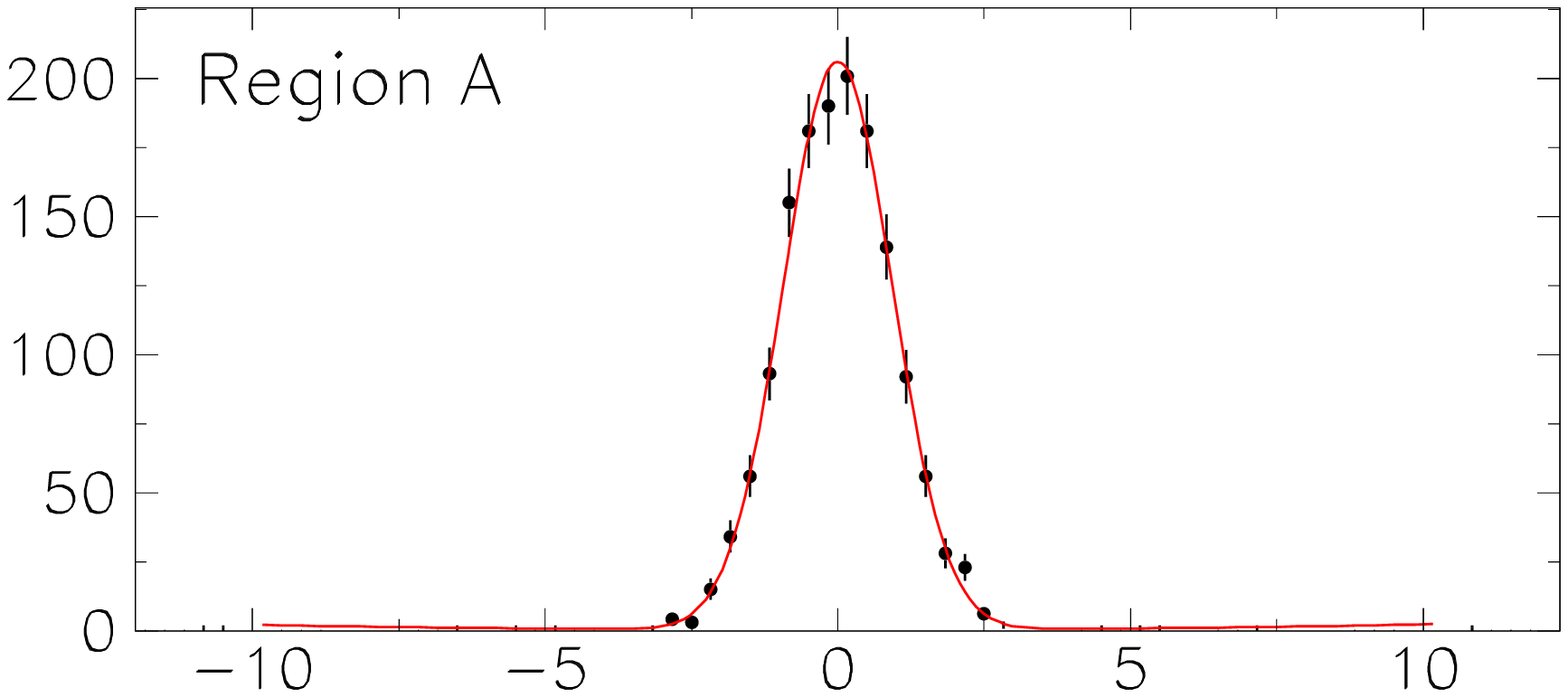}\\
\vglue -0.6cm
\includegraphics[width=7.5cm]{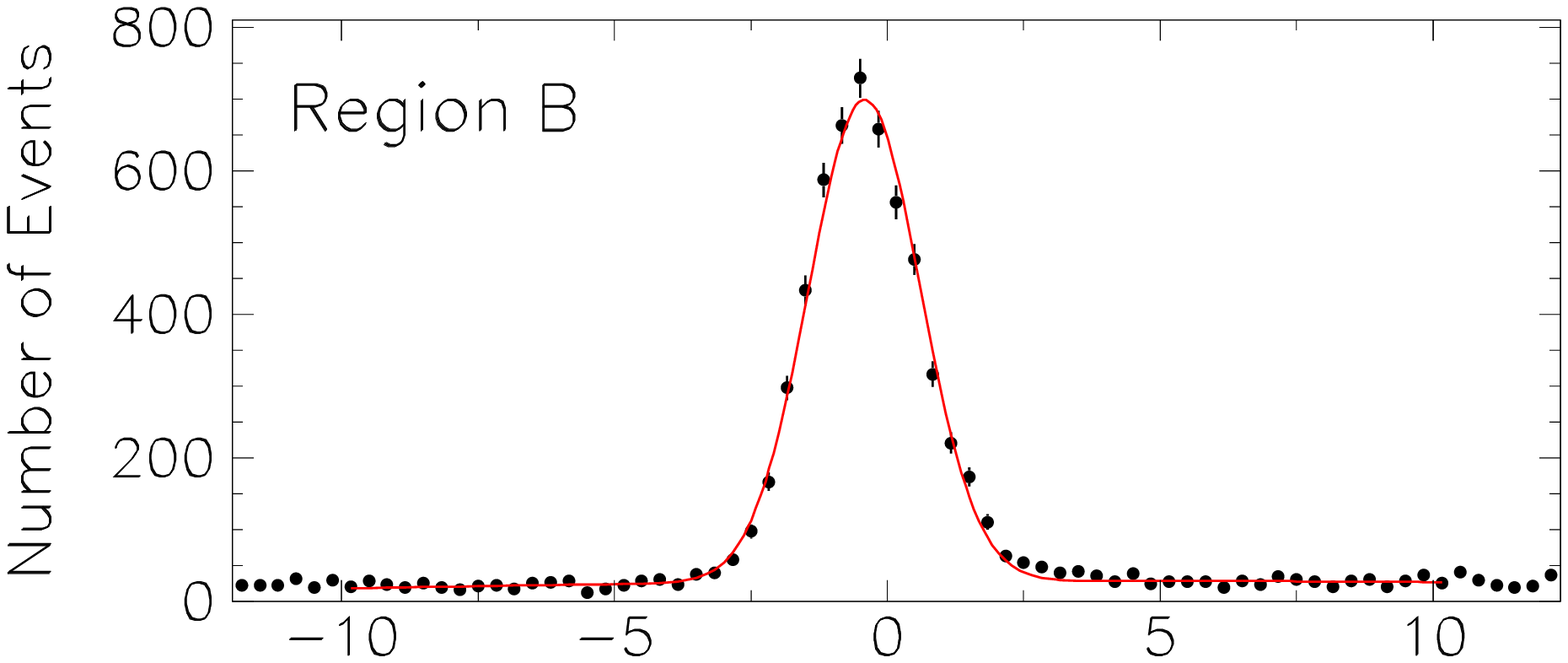}\\
\vglue -0.6cm
\includegraphics[width=7.5cm]{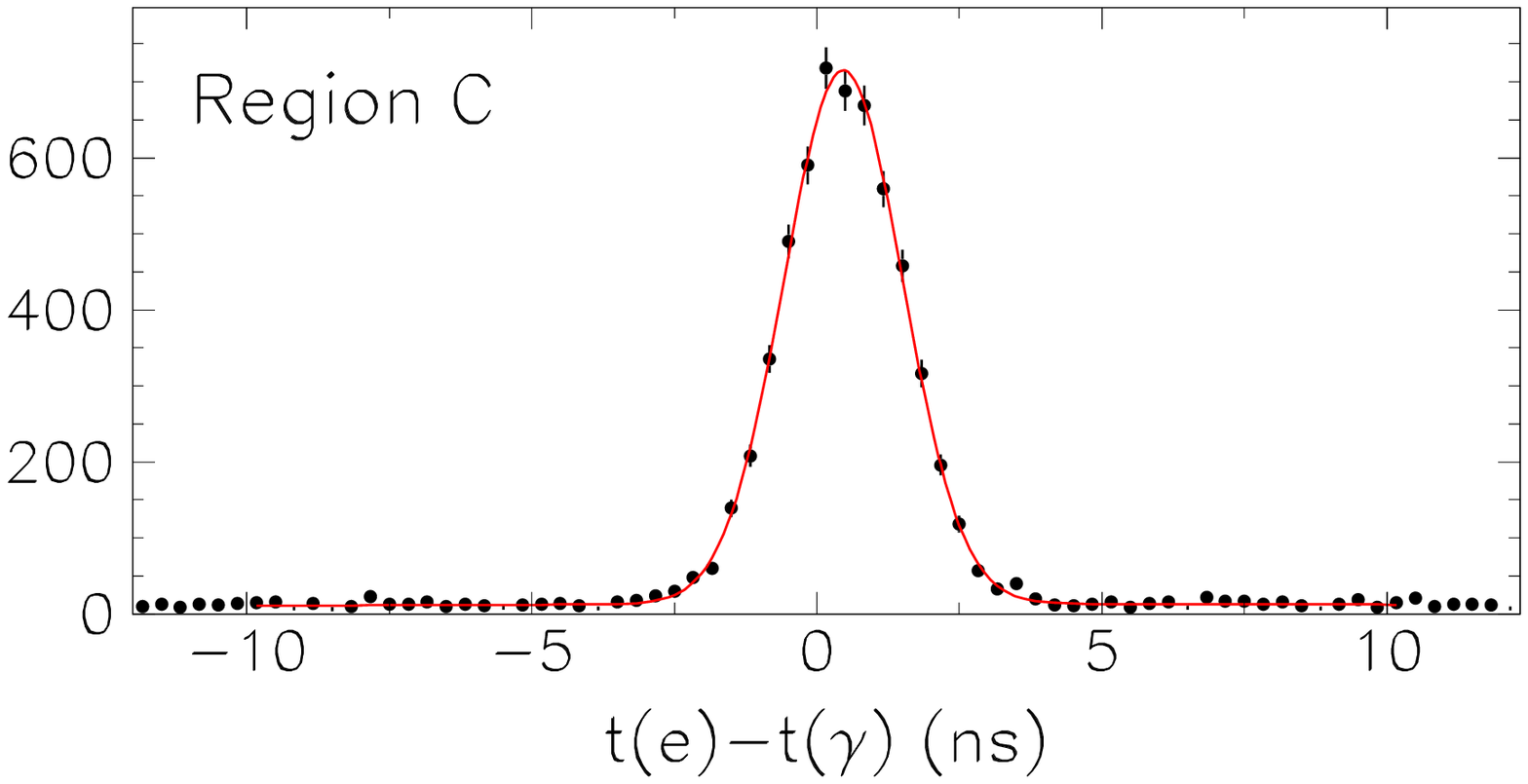}
\vglue -0.5cm
\caption{Signal-to-background ratio for the radiative
pion decay candidate events in three regions of the measured
phase space (see text for details).}
\label{fig:radpi}
\end{figure}
\begin{table*}[ht]
\caption{The fitted and theoretical values of the absolute 
branching ratios for three experimentally accessible regions 
of the phase space in the radiative pion decay.}
\label{tab1}
\newcommand{\m}{\hphantom{$-$}}
\newcommand{\cc}[1]{\multicolumn{1}{c}{#1}}
\renewcommand{\tabcolsep}{2pc} 
\renewcommand{\arraystretch}{1.2} 
\begin{tabular}{@{}lll}
\hline
$E_{e+}^{\rm min}$/$E_{\gamma}^{\rm min}$/$\theta_{e\gamma}^{\rm min}$ & $R^{\rm exp}_{\rm RPD}$ & $R^{\rm the}_{\rm RPD}$ \\
MeV/MeV/deg                                                        & ($\times 10^{-8}$)& ($\times 10^{-8}$) \\
\hline
50/50/$40^\circ$  & 2.655(58) & 2.6410(5) \\
10/50/$40^\circ$  & 14.59(26) & 14.492(5) \\
50/10/$40^\circ$  & 37.95(60) & 37.90(3) \\  
\hline
\end{tabular}\\[2pt]
The fitted $R^{\rm exp}_{\rm RPD}$ values are based on 2004 data set.
\end{table*}

The radiative pion decay $\pi^+\to e^+\nu\gamma$ contributes to
the background of the $\pi\beta$ process, but is also 
interesting in its own right. Precise measurement of its absolute 
branching ratio provides a consistency check of the data 
analysis, and new values of the weak axial and vector $\pi^+$ form factors, 
together with limits on the non-($V-A$) contributions to
Standard Model Lagrangian~\cite{Frl04b}.

We have recorded the radiative pion events in three overlapping
phase space regions:
(1) region $A$, restricted to $e^+$-$\gamma$ coincident pairs for 
which both measured energies in the calorimeter were $E^C_{e^+,\gamma} 
> 55.6\,$MeV, and for which the opening angle was 
$\theta^C_{e^+\gamma} >40.0^\circ$ (3.0\,k events in 2004),
(2) region $B$, with measured positron calorimeter energy $E^C_{e^+} 
> 20.0\,$MeV, the photon energy $E^C_{\gamma} > 55.6\,$MeV and 
the relative angle $\theta^C_{e^+\gamma} >40.0^\circ$ (6.9\,k events), and
(3) region $C$, with measured photon calorimeter energy $E^C_\gamma 
> 20.0\,$MeV, the positron energy $E^C_{e^+} > 55.6\,$MeV and the relative 
angle $\theta^C_{e^+\gamma} >40.0^\circ$ (9.1\,k events).

The signal-to-background timing spectra for all three regions are
shown in Fig.~\ref{fig:radpi}.

In order to account properly for the detector energy-angle resolutions, 
the experimental partial branching ratios were calculated for 
larger regions limited by the physical ``thrown'' kinematic
variables as shown in Table~\ref{tab1}. The integrated radiative corrections
of $-1.0\,$\% (region $A$),  $-1.4\,$\% ($B$), and $-3.3\,$\% ($C$) 
have been added to the theoretical $R^{\rm the}_{\rm RPD}$~\cite{Kur03}. 

The three-dimensional least chi-square ($\chi^2$) fit resulted in a new
experimental value of the weak vector form factor:
\begin{equation}
\label{eq3}
F_V(q^2=0)=0.0262\pm0.0015,
\end{equation}
and the improved value of the weak axial form factor:
\begin{equation}
\label{eq4}
F_A(q^2=0)=0.0118\pm0.0003,
\end{equation}
where $q^2$ stands for the momentum transfer to the lepton pair. 
The third fit parameter was the first measurement of the form
factor's momentum dependence:
\begin{equation}
\label{eq5}
F(q^2)=F(0)\left[ 1+ (0.241\pm0.093)\right]\cdot q^2.
\end{equation}
The above-quoted fit had $\chi^2$ per degree of freedom of 0.6.
The addition of a hypothetical tensor interaction term to the decay 
amplitude (see ~\cite{Bol90,Pob90,Pob92}) results in the upper limit of
$\vert F_T(0)\vert \le 5.1\cdot 10^{-4}$ at the 90\,\% confidence limit~\cite{Max05}.
This limit is more than an order of magnitude smaller than
the ISTRA collaboration re-analysis result reported by Poblaguev~\cite{Pob03}.

\section{RADIATIVE MUON DECAY}
\begin{figure}
\includegraphics[width=7.5cm]{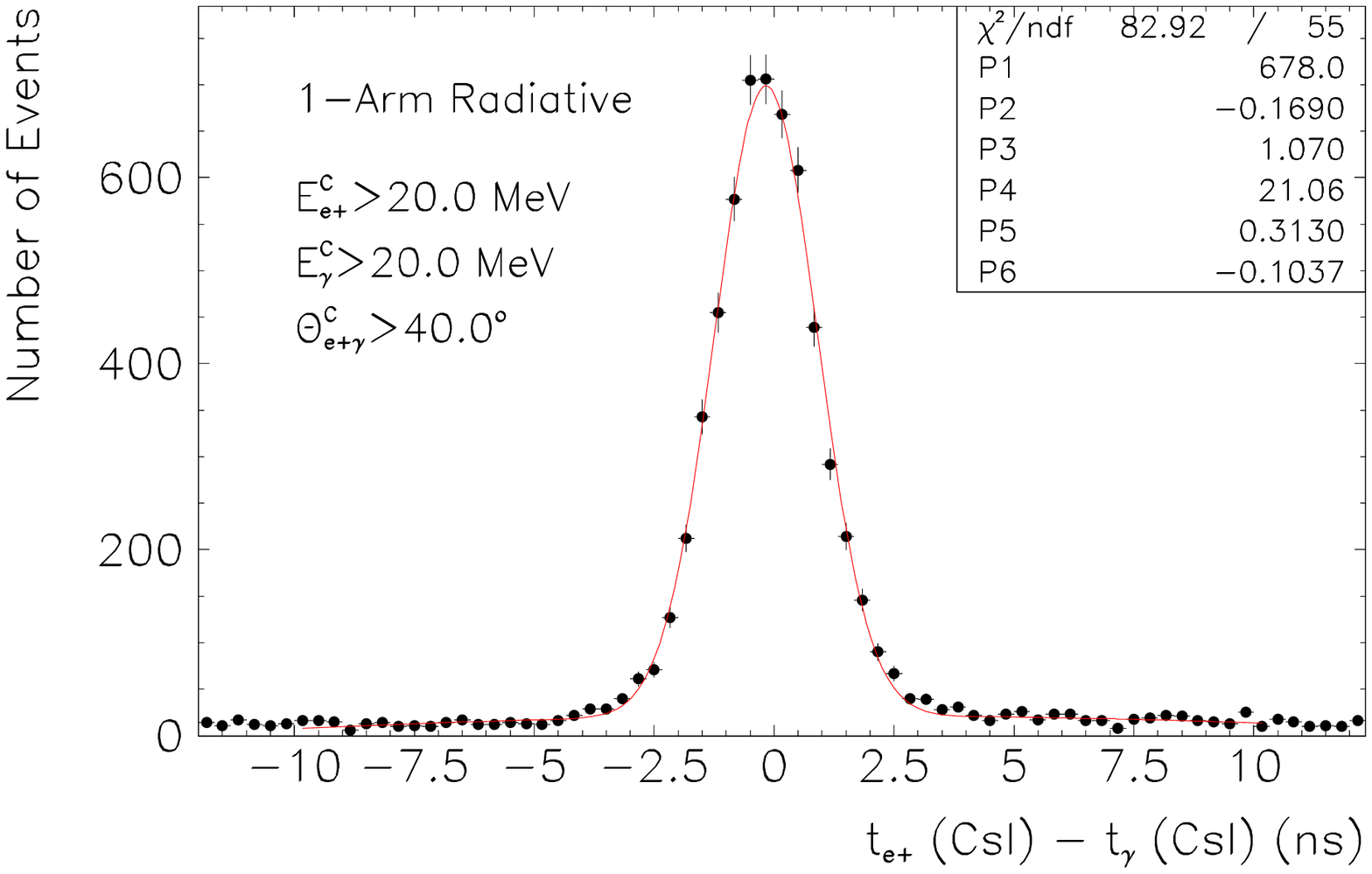}\\
\vglue -1.4cm
\includegraphics[width=7.5cm]{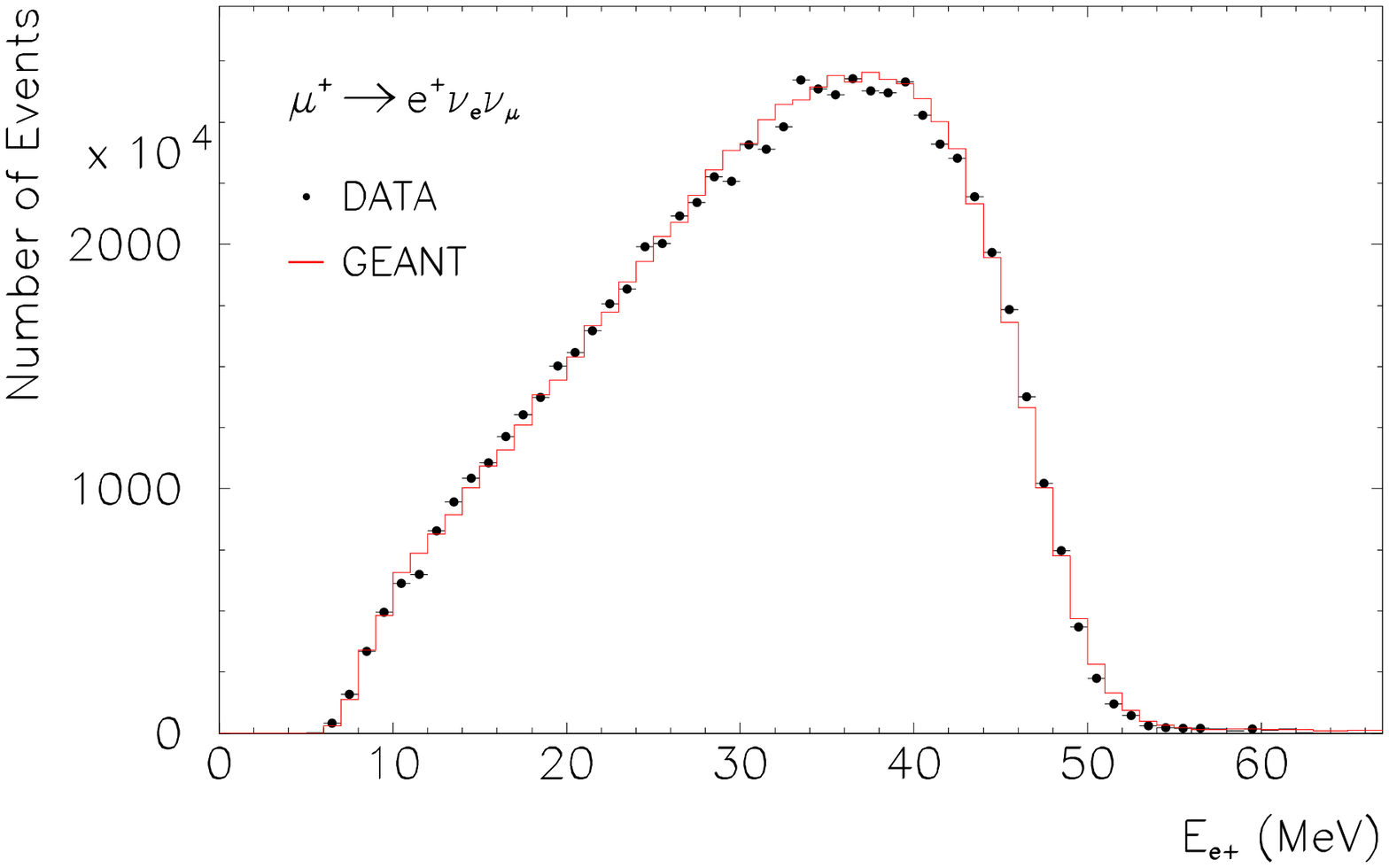}
\vglue -2.0cm
\caption{Signal-to-background ratio for the radiative
muon decay candidate events (top) and energy spectrum
of the normalizing Michel decay (bottom).}
\label{fig:radmu}
\end{figure}
\begin{table*}[htb]
\caption{The optimal values of parameters $\rho$ and $\bar{\eta}$
in the radiative muon decay: two-dimensional fit (the first line)
and the fit with $\rho$ fixed at the Standard Model value 
(the second line).}
\label{tab2}
\newcommand{\m}{\hphantom{$-$}}
\newcommand{\cc}[1]{\multicolumn{1}{c}{#1}}
\renewcommand{\tabcolsep}{2pc} 
\renewcommand{\arraystretch}{1.2} 
\begin{tabular}{@{}ll}
\hline
$\bar{\eta}$ & $\rho$ \\
\hline
$-0.081\pm 0.054{\rm (stat.)}\pm 0.034{\rm (syst.)}$ & $0.751\pm 0.010$ \\
$-0.084\pm 0.050{\rm (stat.)}\pm 0.034{\rm (syst.)}$ & 0.75 (fixed at SM) \\
\hline
\end{tabular}\\[2pt]
The fitted $R^{\rm exp}_{\rm RMD}$ values are based on 2004 data set.
\end{table*}

The radiative muon decay $\mu^+\to e^+\nu\nu\gamma$ measurement
provides another critical consistency check of overall analysis. In the 
Standard Model this process is parameterized via Michel
parameters all of which, save $\bar{\eta}$, can be determined
from the ordinary muon decay~\cite{Mur85}. A non-zero value of $\bar{\eta}$
would imply the non-($V-A$) structure of the electroweak interaction.

The most recent direct measurement of $\bar{\eta}$ can be interpreted
as an upper limit of 0.141 (at 90\,\% CL)~\cite{Eic84}.

Our two-dimensional Michel parameter fit ($\rho,\bar{\eta}$) of
the $4.2\cdot 10^5$ radiative muon events collected in 2004 (Fig.~\ref{fig:radmu})
corresponds to the upper limit $\bar{\eta}\le 0.060$ and simultaneously
yields the SM value $\rho=0.751\pm 0.010$~\cite{Bre06}.
The details are summarized in Table~\ref{tab2}.

\section{RARE {\boldmath$\pi\to e\nu$} DECAY}
\begin{figure}
\includegraphics[width=7.5cm]{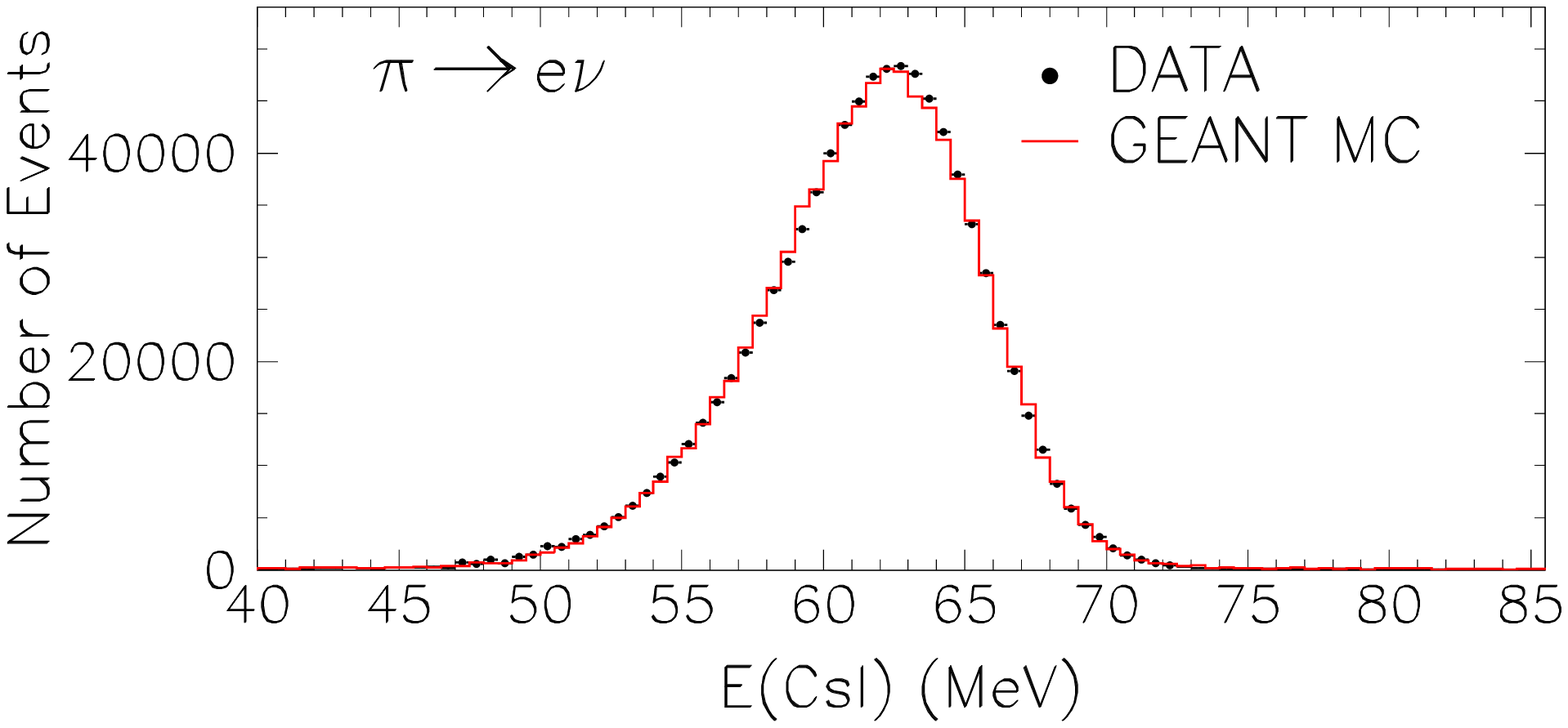}\\
\includegraphics[width=7.5cm]{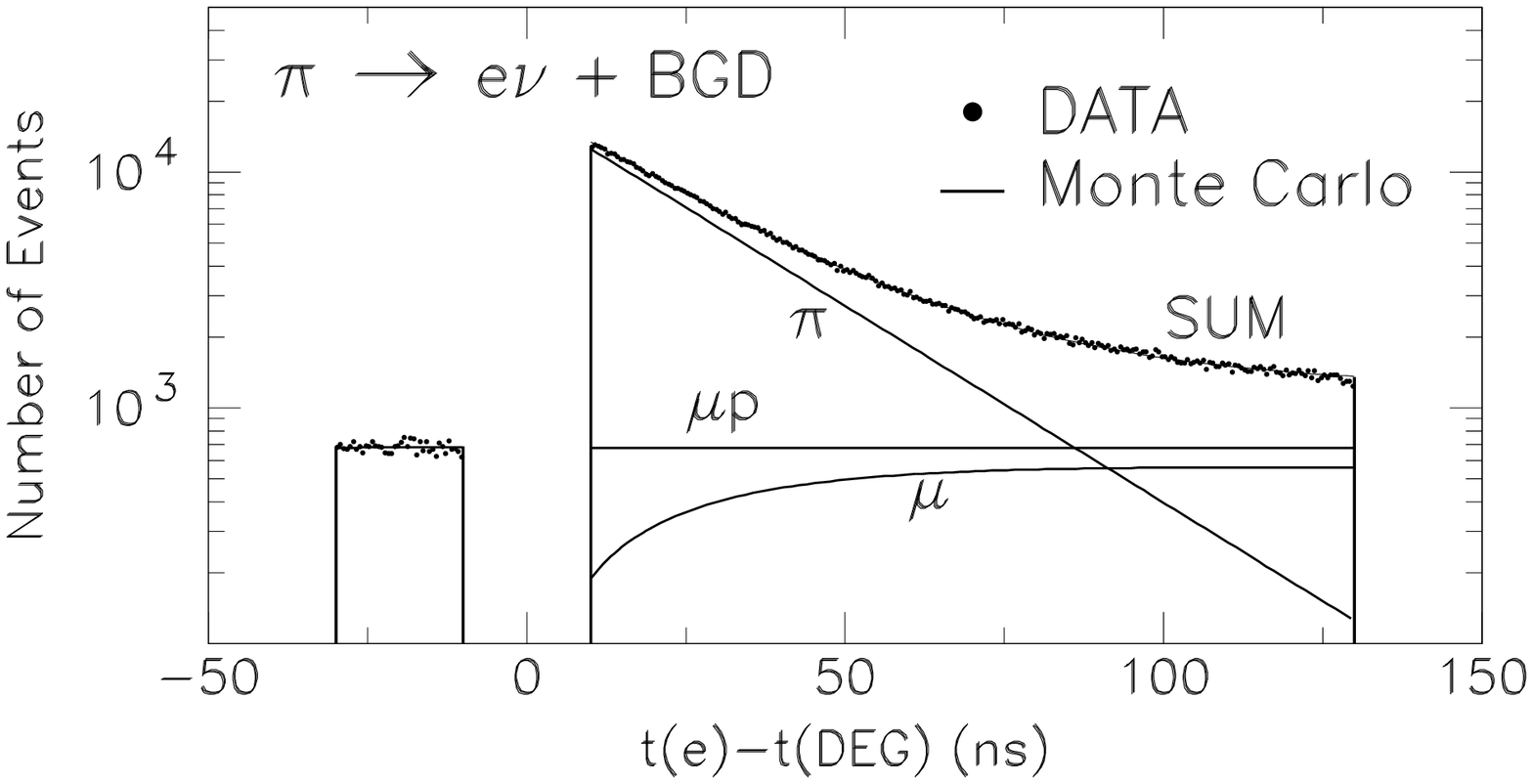}
\vglue -1.0cm
\caption{The energy line shape of the rare 
$\pi^+\to e^+\nu$ events in CsI calorimeter (top) and
timing distribution fit to $\pi^+\to e^+\nu$ candidate
events (bottom).}
\label{fig:p2e}
\end{figure}

We have proposed to perform a new precise measurement of
the $\pi^+\to e^+\nu$ branching ratio at PSI using 
a suitably upgraded PIBETA detector system. The experiment 
has been approved with high priority in 2006. 
The measurement is motivated by the fact that at present
accuracy of that branching ratio lags behind the theoretical
precision by an order of magnitude. 

We will build on our past experience in exploiting the
$\pi\to e\nu$ decay for normalizing purposes (Fig.~\ref{fig:p2e}). 
We plan to rebuild the target region of the detector and
develop the new digitizing detector for the beam counters.
The proposed accuracy of the new measurement is 
$\sim 5\cdot 10^{-4}$ or lower. 

\section{CONCLUSION}

We have reported new and improved absolute branching ratios
for the following rare decays:
(1) $\pi^+\to \pi^0 e^+\nu$, (2) $\pi^+\to e^+\nu\gamma$, and
(3) $\mu^+\to e^+\nu\nu\gamma$. The yields of $\pi^+\to e^+\nu$ 
and $\mu^+\to e^+\nu\nu$ decays that were used for normalization
are also internally consistent when compared to the total 
measured number of decaying $\pi^+$'s and $\mu^+$'s~\cite{Frl04c}.

Our results confirm the CVC hypothesis in the $\pi^+$ system 
at 0.55\,\% level, rule out the tensor contribution in
the radiative $\pi^+$ decay with the form factor 
$\vert F_T\vert \ge 5.1\cdot10^{-4}$
(90\,\% CL), and set the new 90\,\% CL limit on the parameter 
$\bar{\eta}\le 0.060$ in the radiative $\mu^+$ decay.


\begin{thebibliography}{9}
\bibitem{www}
PIBETA home page: http://pibeta.\-phys.\-vir\-gin\-ia.\-edu. 
\bibitem{Frl04a} 
E. Frle\v{z}, D. Po\v{c}ani\'c, K.A. Assamagan et al.,
Nucl. Inst. and Meth. A 526 (2004) 300.
\bibitem{Fey58}
R. P. Feynman and M. Gell-Mann, Phys. Rev. 109 (1958) 193.
\bibitem{Cab63}
N. Cabibbo, Phys. Rev. Lett. 10 (1963) 531.
\bibitem{Jau01}
W. Jaus, Phys. Rev. D 63 (2001) 053009.
\bibitem{Cir03}
V. Cirigliano, M. Knecht, H. Neufeld, and H. Pichl,
Eur. Phys. J. C 27 (2003) 255.
\bibitem{McF85}
W.K. McFarlane, L.B. Auerbach, F.C. Gaille et al.,
Phys. Rev. D 32 (1985) 547. 
\bibitem{Mar86}
W. J. Marciano and A. Sirlin,  Phys. Rev. Lett. 56 (1986) 22. 
\bibitem{Bri92}
D.I. Britton, S. Ahmad, D.A. Bryman at al.,
Phys. Rev. Lett. 68 (1992) 3000.
\bibitem{Cza93}
G. Czapek, A. Federspiel, A. Fl\"ukiger et al.,
Phys. Rev. Lett. 70 (1993) 17.
\bibitem{Mar93} 
W.J. Marciano, A. Sirlin, Phys. Rev. Lett. 71 (1993) 3629.
\bibitem{Dec95}
R. Decker and M. Finkemeier, Nucl. Phys. B 438 (1995) 17.
\bibitem{Eid04}
S. Eidelman et al., Physics Letters B592 (2004) 1.
\bibitem{Poc04} 
D. Po\v{c}ani\'c, E. Frle\v{z}, V.A. Baranov et al.,
Phys. Rev. Lett. 93 (2004) 181803-1.
\bibitem{Frl04b} 
E. Frle\v{z}, D. Po\v{c}ani\'c, V.A. Baranov et al.,
Phys. Rev. Lett. 93 (2004) 181804-1.
\bibitem{Kur03}
E.~A.~Kuraev, Yu.~M.~Bystritsky and E.~P.~Velicheva,
Private communication (2003).
\bibitem{Bol90}
V.~N.~Bolotov, S.~N.~Gninenko, R.~M.~Djilkibaev et al., 
Phys. Lett. B243 (1990) 308.
\bibitem{Pob90}
A.~A.~Poblaguev, Phys. Lett. B238 (1990) 108.
\bibitem{Pob92}
A.~A.~Poblaguev, Phys. Lett. B286 (1992) 169.
\bibitem{Max05}
M.A. Bychkov, Ph. D. Thesis, University of Virginia (2005).
\bibitem{Pob03}
A.A. Poblaguev, Phys. Rev. D 68 (2003) 054020.
\bibitem{Mur85}
K. Mursula and F. Schenk, Nucl. Phys. B253 (1985) 189.
\bibitem{Eic84}
W. Eichenberger, R. Engfer, and A. van~der~Schaaf,
Nucl. Phys.~A, 412 (1984) 523. 
\bibitem{Bre06}
B.A. VanDevender, Ph. D. Thesis, University of Virginia (2006).
\bibitem{Frl04c}
E. Frle\v{z}, Fizika B 13 (2004) 243. 

\end{thebibliography}
\end{document}